\documentclass{ws-procs9x6}
\usepackage{epsfig}
\begin{document}

\title{Double Transverse Spin Asymmetries at Next-to-leading Order in QCD}

\author{A. MUKHERJEE}
\address{Lorentz Institute, University of Leiden, 2300 RA Leiden, The Netherlands}

\author{M. STRATMANN}

\address{Institute f\"ur Theoretische Physik, Universit\"at Regensburg \\ 
D 93040 Regensburg, Germany}
\author{W. VOGELSANG}

\address{(1) Physics Department, Brookhaven National Laboratory
\\Upton, New York 11973, USA \\ 
(2) RIKEN-BNL Research Center, Bldg. 510a, Brookhaven National Laboratory,\\
Upton, New York 11973 -- 5000, USA}

\address{\vspace*{-10cm}\hspace*{9cm}
BNL-NT-04/39 \\[2mm] \hspace*{9.5cm}
RBRC-477 \\[8.6cm]}

\maketitle

\abstracts{We present a technique to calculate the cross sections and spin
asymmetries for transversely polarized $pp$ collisions at NLO  in QCD
and report on the use of this technique for the processes $p^{\uparrow}
 p^{\uparrow} \rightarrow \gamma X$, $p^{\uparrow}
p^{\uparrow} \rightarrow \pi X$ and  $p^{\uparrow}
p^{\uparrow} \rightarrow l^+ l^- X$.}  

\section{Introduction}
Combined experimental and theoretical efforts in the past few years 
have led to an improved
understanding of the unpolarized parton distributions $f(x,Q^2)$ and  the
helicity distributions $\Delta f(x,Q^2)$ of the nucleon. It is known that the
complete understanding of the partonic structure of a spin ${1\over 2}$
object like a nucleon is given in terms of $f(x,Q^2)$, $\Delta f (x, Q^2)$
and by the transversity distributions $\delta f(x, Q^2)$, which give the
number densities of partons having the same polarization as the nucleon,
when the nucleon is transversely polarized, minus the number with opposite
polarization. $\delta f(x,Q^2)$ remain quantities about which we
have the least knowledge and are at present the focus of much experimental
activity.

    Transversity will be probed in the double transverse spin asymmetries in
transversely polarized $pp$ collisions at the BNL Relativistic Heavy Ion
Collider (RHIC). The potential of RHIC in accessing transversity through
double transverse spin asymmetries $A_{TT}$ in the Drell-Yan process was
estimated in \cite{martin}. Other relevant processes include high $p_T$
prompt photon and jet production \cite{lo}. Apart from DY, the other
calculations were done at leading order (LO). It is known that the 
next-to-leading order (NLO) QCD calculations are necessary in order to have
a firm theoretical prediction.

\section{Projection Technique}

Apart from the motivations given above, interesting new technical questions
arise beyond LO in the calculations of cross sections involving transverse
polarization. Unlike the longitudinally polarized case, where the spin
vectors are aligned with the momentum, the transverse spin vectors specify
extra spatial directions and as a result, the cross section has non-trivial
dependence  on the azimuthal angle of the observed particle. For $A_{TT}$
this dependence is always of the form \cite{phi} 
\begin{eqnarray}
\label{eq2}
\frac{d^3\delta \sigma}{dp_T d\eta d\Phi}\;\equiv\;
\cos (2\Phi)\,\left\langle \frac{d^2\delta\sigma}{dp_T d\eta}\right\rangle
\; ,
\end{eqnarray}
for a parity conserving theory with vector coupling, here the $z$ axis is
defined by the direction of the initial partons in their center-of-mass
frame and the spin vectors are taken to point in the $\pm x$ direction.
Therefore the integration over the azimuthal angle is not appropriate. This
makes it difficult to use the standard techniques developed for NLO
calculations of  
unpolarized and longitudinally polarized processes here because all these
techniques usually rely on the integration over the full azimuthal phase
space and also on particular reference frames which are related in a
complicated way to the center-of mass frame of the initial protons. In
\cite{photon} a new general technique was introduced which facilitates NLO
calculations with transverse polarizations by conveniently projecting on the
azimuthal dependence of the cross section in a covariant way. The projector
\begin{eqnarray}
\label{eq5}
F(p,s_a,s_b) =
\frac{s}{\pi t u} 
\,\left[ 2 \,(p\cdot s_a)\, (p\cdot s_b)\; +\;
\frac{t u}{s} \,(s_a \cdot s_b) \right] \;,
\end{eqnarray}
reduces to ${cos 2 \phi\over \pi}$ in the center-of-mass frame of the
initial protons. Here $p$ is the momentum of the observed particle in the
final state. The cross section is multiplied  with the projector and          
integrated over the full azimuthal phase space. Integrations of the terms
involving the product of the transverse spin vectors $s_a, s_b$ with the
momenta can be performed using a tensor decomposition. After this step, there
are no scalar products involving $s_i$ left in the matrix element. For the
integration over the phase space, one can now use the standard techniques
from the unpolarized and longitudinally polarized cases. This method is
particularly convenient at NLO, where one uses dimensional regularization and
the phase space integrations are performed in $n$ dimensions.

\section{Applications}
As an example, we discuss the use of this technique for high $p_T$ prompt
photon production. The LO process is $q {\bar q} \rightarrow \gamma g$. 
We multiply $\delta |M|^2$ by the projector $ F(p,s_a,s_b) $ and integrate
over the full $\Delta \phi$ in a covariant way. At NLO, there are two
subprocesses contributing, $qq \rightarrow \gamma X $, where $X = qq$ and $q
{\bar q} \rightarrow \gamma X $, where $X = q{\bar q}+gg+q'{\bar q'}$. For
$2 \rightarrow 3$ processes, one integrates over the  phase spaces of the
unobserved particles, after multiplying with the projector and
eliminating the scalar products with the spin vectors using tensor
decomposition. Owing to the presence of the ultraviolet, infrared and
collinear singularities, one has to introduce a regulator. We choose
dimensional regularization. UV poles in the virtual diagrams are removed by
renormalization of the strong coupling constant. Infrared singularities
cancel between the real emission and virtual diagrams. After this, only
collinear singularities remain, which result from collinear splitting of an
initial-state parton into a pair of partons.  These correspond to
long-distance contribution to the partonic cross section. From the
factorization theorem it follows that such contributions need to be factored
into the parton distributions. We have imposed an isolation cut \cite{frix}       
to remove the background contribution. All final-state collinear
singularities then cancel. The isolation constraint was imposed analytically
by assuming a narrow isolation cone.

For our numerical predictions, we model the transversity distribution by
saturating Soffer's inequality \cite{soffer} at some low input scale
$\mu_0 \approx 0.6 $ GeV and for higher scales, $\delta f(x, \mu)$  are
obtained by solving the QCD evolution equations. Figure 1 shows the results
for the prompt photon production in transversely polarized $pp$ collisions.
Our numerics apply to the PHENIX detector at RHIC. The lower part of the
figure displays the '$K$- factor', $ K=\frac{d\delta\sigma^{\rm
NLO}}{d\delta\sigma^{\rm LO}}$. The scale dependence becomes much weaker at
NLO. The corresponding asymmetries are given in \cite{photon}.        

\begin{figure}[th]
\vspace*{-0.5cm}  
\begin{center}    
\epsfig{figure=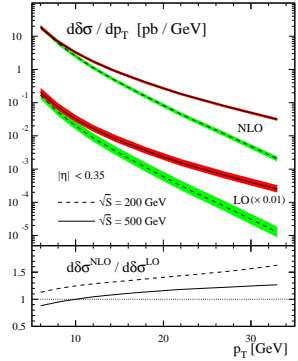,width=0.4\textwidth}
\end{center}
\vspace*{-0.5cm}
\caption{\sf Predictions for the transversely polarized prompt
photon production cross sections at LO and NLO, for $\sqrt{S}=200$
and 500 GeV. The LO results have been scaled by a factor of 0.01. 
The shaded bands represent the theoretical uncertainty if
$\mu_F$ $(=\mu_R)$ is varied in the range $p_T/2\le \mu_F \le 2p_T$.
The lower panel shows the ratios of the NLO and LO results for both 
c.m.s.\ energies. 
\label{fig:sigma}}
\end{figure}

For the Drell-Yan lepton pair production in transversly polarized $pp$
collisions the LO subprocess is $q {\bar q} \rightarrow l^+l^-$. The real
emission $2 \rightarrow 3$ subprocess is  $q {\bar q} \rightarrow l^+l^-g$.
We multiply the squared matrix element by the projector and integrate over
the phase space. We obtain the known result for the DY coefficient function
in the $\overline {MS}$ scheme at NLO \cite{photon}. 

For inclusive pion production in transversely polarized process the LO
channels are $qq \rightarrow qX, q{\bar q} \rightarrow qX,
q{\bar q} \rightarrow q'X, q{\bar q} \rightarrow gX $. At NLO there are
$O(\alpha_s)$ corrections to the above processes and the additional channel
$qq \rightarrow gX$. We have used the projection technique  to
calculate the cross section at NLO and the numerical results are in 
progress.

\section*{Acknowledgments}
AM thanks the organizers of the $16$ th International Spin Physics Symposium
for a wonderful conference and hospitality and FOM for support.
WV\ is grateful to RIKEN, Brookhaven National Laboratory
and the U.S.\ Department of Energy (contract number DE-AC02-98CH10886) for
providing the facilities essential for the completion of his work.

\end{document}